\documentclass[authoryear,final,5p,times,twocolumn]{elsarticle}

 \usepackage{graphicx}

\usepackage{amssymb}

\usepackage{lscape}

\journal{New Astronomy}

\def\aap{A\&A}
\def\aaps{A\&AS}
\def\aj{AJ}
\def\apj{ApJ}
\def\apjs{ApJS}

\def\mnras{MNRAS}

\def\bain{Bull. Astron. Inst. Netherlands}

\begin{document}

\begin{frontmatter}



\title{Spectroscopic study of the O-type runaway supergiant HD\,195592}


\author{M. De Becker\corref{cor1}}
\ead{debecker@astro.ulg.ac.be}
\cortext[cor1]{Postdosctoral Researcher FRS/FNRS (Belgium)}

\author{N. Linder}
\ead{linder@astro.ulg.ac.be}

\author{G. Rauw}
\ead{rauw@astro.ulg.ac.be}

\address{Institut d'Astrophysique et de G\'eophysique, Universit\'e de Li\`ege, 17, All\'ee du 6 Ao\^ut, B5c, B-4000 Sart Tilman, Belgium}

\begin{abstract}
The scope of this paper is to perform a detailed spectroscopic study of the northern O-type supergiant HD\,195592. We use a large sample of high quality spectra in order to investigate its multiplicity, and to probe the line profile variability. Our analysis reveals a clear spectroscopic binary signature in the profile of the He\,{\sc i}\,$\lambda$\,6678 line, pointing to a probable O + B system. We report on low amplitude radial velocity variations in every strong absorption line in the blue spectrum of HD\,195592. These variations are ruled by two time-scales respectively of 5.063 and about 20 days. The former is firmly established, whilst the latter is poorly constrained. We report also on a very significant line profile variability of the H\,$\beta$ line, with time scales strongly related to those of the radial velocities. Our results provide significant evidence that HD\,195592 is a binary system, with a period that might be the variability time-scale of about 5\,days. The second time scale may be the signature of an additional star moving along a wider orbit provided its mass is low enough, even though direct evidence for the presence of a third star is still lacking. Alternatively, the second time-scale may be the signature of a variability intrinsic to the stellar wind of the primary, potentially related to the stellar rotation.
\end{abstract}


\end{frontmatter}


\section{Introduction \label{intro}}
Most O-type stars are found to be part of open clusters, where they are believed to be formed. However, some stars are characterized by large peculiar space velocities and seem to have been ejected from their birth place. Such objects -- the so-called runaway stars -- are important in the sense that they provide the opportunity to investigate quite unusual events, namely processes at the origin of stellar ejection from open clusters. Mainly, two scenarios have been proposed to explain the peculiar kinematics of runaway stars. On the one hand, one may consider a supernova explosion in a close binary (or higher multiplicity) system, resulting in the ejection of the secondary \citep{zwicky,blaauw}. In the context of this scenario, the resulting neutron star may remain bound to the secondary. According to \citet{portegiesrunaway}, about 20 -- 40\,$\%$ of the runaways may indeed be binaries harbouring a compact companion. On the other hand, the ejection of the star can be the result of dynamical interactions in a dense open cluster \citep[see e.g.][]{leonardduncan}. In the latter case, a binary fraction of about 10\,$\%$ is expected. Multiplicity studies of runaway stars are needed in order to check whether the dynamical interaction scenario is sufficient to explain the observed binary fraction, or if the supernova explosion scenario is needed.

\begin{table}
\begin{center}
\caption{Observing runs. The first and second columns give the name of the campaign as used in the text. The next columns yield the number of spectra obtained, the time elapsed between the first and the last spectrum of the run ($\Delta$T), the natural width of a peak of the power spectrum taken as 1/$\Delta$T, and finally the mean signal-to-noise ratio of each data set.\label{runs}}
\begin{tabular}{l c c c c}
\hline\hline
 Obs. run &  N & $\Delta$T & $\Delta\nu_\mathrm{nat}$ & S/N \\
  &   & (d) & (d$^{-1}$) & \\
\hline
\vspace*{-0.2cm}\\
Oct.2004 & 4 & 9.2 & 0.11 & 150 \\
Sum.2005 & 7 & 8.0 & 0.13 & 450 \\
Sept.2006 & 20 & 4.2 & 0.24 & 450 \\
Oct.2006 & 2 & 2.0 & 0.50 & 380 \\
Aut.2007 & 48 & 27.0 & 0.04 & 500 \\
\vspace*{-0.2cm}\\
\hline
\end{tabular}
\end{center}
\end{table}

\begin{table*}
\begin{center}
\caption{Journal of the observations of HD\,195592. The heliocentric Julian date is given as HJD - 2\,450\,000, and the date (yyyy/mm/dd) is that of the beginning of the night. For each observation, the mean radial velocity expressed in km\,s$^{-1}$ is given. The lines used to computed the mean radial velocity are He\,{\sc i} $\lambda$ 4471, Si\,{\sc iii} $\lambda\lambda$ 4552,4567, C\,{\sc iii} $\lambda\lambda$ 4647,4650, and He\,{\sc i} $\lambda$ 4713. \label{log}}
\begin{tabular}{ l c c c | l c c c | l c c c}
\hline\hline
\vspace*{-0.2cm}\\
No. & HJD  & Date & RV & No. & HJD  & Date & RV & No. & HJD  & Date & RV \\
\vspace*{-0.2cm}\\
\hline
1 & 3286.301 & 2004/10/07 & --36.0 & 28 & 3983.478 & 2006/09/04 & --30.2 & 55 & 4410.244 & 2007/11/05 & --20.4 \\
2 & 3287.386 & 2004/10/08 & --28.2 & 29 & 3984.310 & 2006/09/05 & --29.6 & 56 & 4410.321 & 2007/11/05 & --24.0 \\
3 & 3290.450 & 2004/10/11 & --26.3 & 30 & 3984.390 & 2006/09/05 & --31.2 & 57 & 4411.251 & 2007/11/06 & --37.3 \\
4 & 3295.455 & 2004/10/16 & --21.0 & 31 & 3984.465 & 2006/09/05 & --33.6 & 58 & 4411.283 & 2007/11/06 & --30.6 \\
5 & 3547.578 & 2005/06/25 & --16.1 & 32 & 4033.258 & 2006/10/24 & --21.5 & 59 & 4411.314 & 2007/11/06 & --30.8 \\
6 & 3548.548 & 2005/06/26 & --31.5 & 33 & 4035.255 & 2006/10/26 & --33.1 & 60 & 4412.242 & 2007/11/07 & --16.7 \\
7 & 3550.573 & 2005/06/28 & --26.7 & 34 & 4396.268 & 2007/10/22 & --34.8 & 61 & 4412.267 & 2007/11/07 & --21.4 \\
8 & 3551.533 & 2005/06/29 & --33.2 & 35 & 4396.343 & 2007/10/22 & --31.3 & 62 & 4413.242 & 2007/11/08 & --20.8 \\
9 & 3553.548 & 2005/07/01 & --26.6 & 36 & 4397.232 & 2007/10/23 & --21.0 & 63 & 4413.330 & 2007/11/08 & --23.5 \\
10 & 3554.598 & 2005/07/02 & --21.5 & 37 & 4397.290 & 2007/10/23 & --21.0 & 64 & 4414.264 & 2007/11/09 & --28.5 \\
11 & 3555.555 & 2005/07/03 & --43.6 & 38 & 4400.296 & 2007/10/26 & --33.9 & 65 & 4414.368 & 2007/11/09 & --25.4 \\
12 & 3980.302 & 2006/09/01 & --32.4 & 39 & 4401.243 & 2007/10/27 & --29.8 & 66 & 4415.246 & 2007/11/10 & --37.6 \\
13 & 3980.371 & 2006/09/01 & --33.7 & 40 & 4401.295 & 2007/10/27 & --24.2 & 67 & 4415.333 & 2007/11/10 & --33.4 \\
14 & 3980.437 & 2006/09/01 & --35.4 & 41 & 4402.235 & 2007/10/28 & --26.0 & 68 & 4415.363 & 2007/11/10 & --39.2 \\
15 & 3980.512 & 2006/09/01 & --35.4 & 42 & 4402.289 & 2007/10/28 & --19.8 & 69 & 4416.228 & 2007/11/11 & --33.0 \\
16 & 3981.366 & 2006/09/02 & --36.8 & 43 & 4403.276 & 2007/10/29 & --18.5 & 70 & 4416.301 & 2007/11/11 & --27.9 \\
17 & 3981.442 & 2006/09/02 & --33.9 & 44 & 4403.343 & 2007/10/29 & --17.7 & 71 & 4417.223 & 2007/11/12 & --36.1 \\
18 & 3982.306 & 2006/09/03 & --23.3 & 45 & 4405.245 & 2007/10/31 & --28.7 & 72 & 4417.248 & 2007/11/12 & --37.7 \\
19 & 3982.347 & 2006/09/03 & --23.5 & 46 & 4405.302 & 2007/10/31 & --29.9 & 73 & 4418.223 & 2007/11/13 & --29.6 \\
20 & 3982.413 & 2006/09/03 & --21.6 & 47 & 4406.236 & 2007/11/01 & --20.8 & 74 & 4419.225 & 2007/11/14 & --31.3 \\
21 & 3982.440 & 2006/09/03 & --24.2 & 48 & 4406.291 & 2007/11/01 & --18.9 & 75 & 4419.250 & 2007/11/14 & --31.5 \\
22 & 3982.496 & 2006/09/03 & --23.7 & 49 & 4407.244 & 2007/11/02 & --20.3 & 76 & 4421.221 & 2007/11/16 & --34.3 \\
23 & 3983.330 & 2006/09/04 & --26.4 & 50 & 4407.324 & 2007/11/02 & --23.9 & 77 & 4421.305 & 2007/11/16 & --30.5 \\
24 & 3983.356 & 2006/09/04 & --28.2 & 51 & 4408.242 & 2007/11/03 & --15.5 & 78 & 4422.230 & 2007/11/17 & --15.2 \\
25 & 3983.376 & 2006/09/04 & --28.2 & 52 & 4408.320 & 2007/11/03 & --11.9 & 79 & 4422.327 & 2007/11/17 & --13.4 \\
26 & 3983.421 & 2006/09/04 & --27.8 & 53 & 4409.242 & 2007/11/04 & --20.2 & 80 & 4423.225 & 2007/11/18 & --20.5 \\
27 & 3983.459 & 2006/09/04 & --29.1 & 54 & 4409.318 & 2007/11/04 & --19.2 & 81 & 4423.292 & 2007/11/18 & --23.6 \\
\vspace*{-0.2cm}\\
\hline
\end{tabular}
\end{center}
\end{table*}

In a recent spectroscopic study, \citet{mcswainrunaway} investigated a series of field OB stars in order to check their runaway status. They discovered that seven of them have been ejected from their birth place. Three of them were also reported to be associated with bow shocks due to their interaction with the interstellar medium \citep{bowshockrunaway}. Among these runaways, one finds HD\,195592 (O9.5Ia) to which this paper is devoted. The study of \citet{masonspeckle} did not reveal any visual companion, whilst  \citet{plaskettbin} suggested it may be a spectroscopic binary. \citet{mcswainrunaway} reported also on radial velocity variations but the binary status of HD\,195592 still needed to be confirmed. Our target star is located at a distance of about 1.1\,kpc, and may be originating from the open cluster NGC\,6913 \citep{SchilRoser}.

In this paper, we report on an intensive multiplicity study of HD\,195592 based on a high quality time series of spectra obtained in the blue domain. We first present our analysis, including radial velocity measurements and variability tests (Section\,\ref{res}). We then discuss the results and present our conclusions respectively in Sections\,\ref{disc} and \ref{conc}.

\section{Observations and data reduction}\label{obs}
Spectroscopic observations were collected at the Observatoire de Haute-Provence (OHP, France) during several observing runs from October 2004 to November 2007. All spectra were obtained with the Aur\'elie spectrograph fed by the 1.52\,m telescope \citep{aurelie}. The typical exposure time was about 20 -- 30 minutes, depending on the weather conditions. Our collection of blue spectra (between 4460 and 4890\,\AA\,) is described in Table\,\ref{runs}. All our data have been treated following the procedure described by \citet{ic1805_1}. The signal-to-noise ratio of our spectra was estimated in a region devoid of spectral lines, between 4790 and 4800\,\AA\,. We note that the quality of the spectra is rather uniform within each observing run, with a dispersion in signal-to-noise ratio lower than about 20\,\% with respect to the mean value given in Table\,\ref{runs}.

The complete data set in the blue contains 81 spectra, spread over a time interval of about 3 years. We managed the observational strategy to cover many time-scales, from a few hours up to a few years. In 2006, we obtained indeed several spectra per night during a few nights to cover short time-scales. In the 2007 observing run, we obtained generally 2 spectra per night during almost 28 consecutive nights. The complete journal of observations is given in Table\,\ref{log}.

In addition, we also obtained 4 spectra in the red domain (between 6340 and 6780\,\AA\,) using the same setup. In this spectral domain, the resolving power is of the order of 11000. The HJD -- 2\,450\,000 at mid-exposure are respectively 3549.571, 3555.377, 3555.523, and 3652.417. 

\section{Results \label{res}}

\begin{figure*}
\centering
\includegraphics[width=170mm]{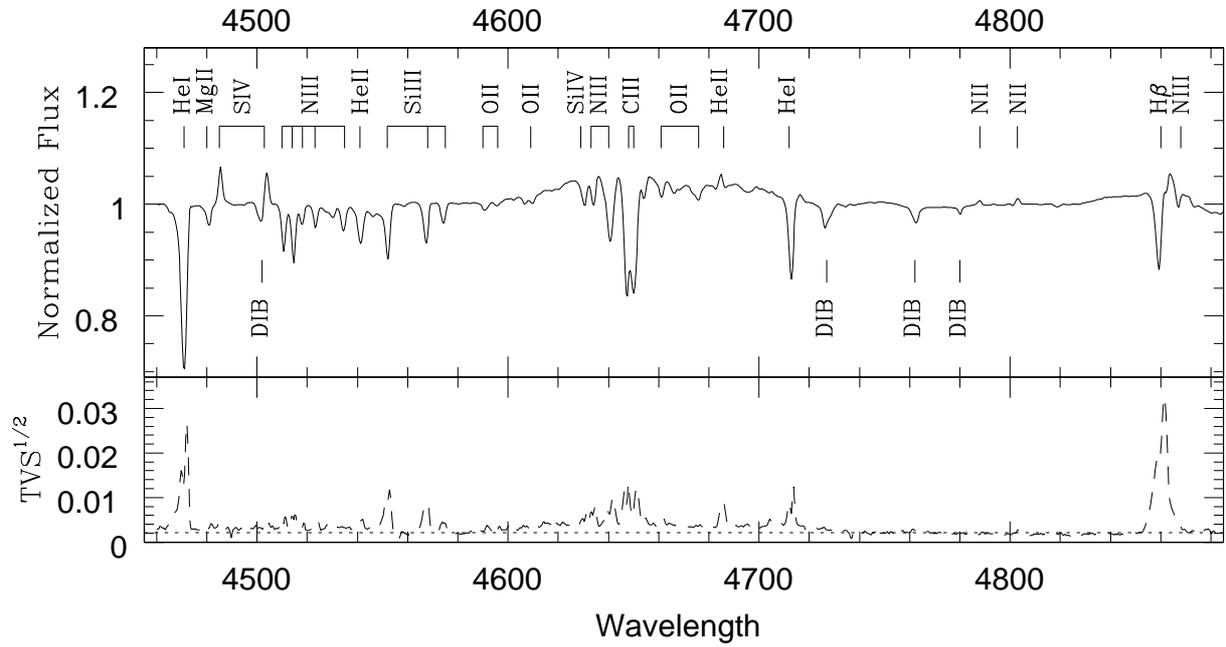}
\caption{Blue spectrum of HD\,195592 between 4460 and 4890 \AA\,. {Upper part:} mean spectrum with line identifications. {\it Lower part:} TVS with its 99\,$\%$ confidence level illustrated by the horizontal dashed line.}
\label{spectvs}
\end{figure*}

\begin{figure*}
\centering
\includegraphics[width=170mm]{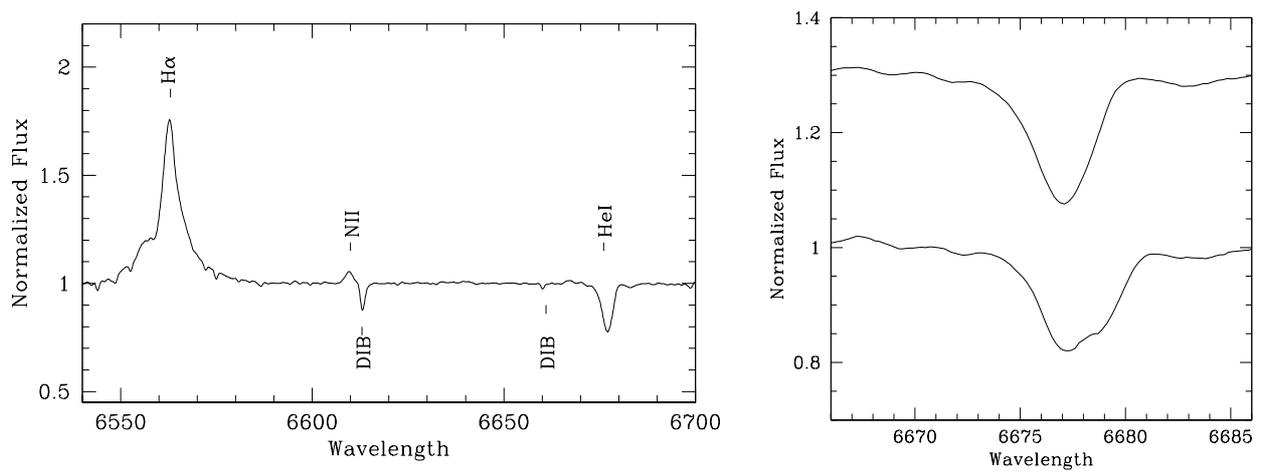}
\caption{{\it Left part:} Red spectrum of HD\,195592 between 6540 and 6700 \AA\,. {\it Right part:} Line profile of He\,{\sc i} $\lambda$ 6678\,\AA\, respectively at HJD\,2\,453\,459.571 (top) and HJD\,2\,453\,652.417 (bottom). The line shape of the lower profile strongly suggests a partial deblending of two spectral lines.}
\label{specred}
\end{figure*}

The mean blue spectrum of HD\,195592, i.e. between 4460 and 4890\,\AA\,, with line identifications\footnote{We note the presence of two weak unusual emission lines respectively at 4788 and 4803\,\AA\,. We attribute them to N\,{\sc ii} \citep[see e.g.][]{kay}.} is shown in the upper panel of Figure\,\ref{spectvs}. We first searched for a significant variability in the spectral lines using the Temporal Variance Spectrum (TVS) analysis method developed by \citet{fullertontvs}. In the lower panel of Figure\,\ref{spectvs}, we show that we detect a significant variability (above the 99\,$\%$ confidence level) for most of the lines, with a higher TVS amplitude for the strongest lines. On the other hand, the red spectrum of HD\,195592 is dominated by the emission of H\,$\alpha$ (left part of Figure\,\ref{specred}), and presents also N\,{\sc ii}\,$\lambda$\,6610 in emission and He\,{\sc i}\,$\lambda$\,6678 in absorption. The inspection of our sample of red spectra revealed immediately an asymmetric profile for the latter line (right part of Figure\,\ref{specred}), suggesting a partial deblending of two lines reminiscent of the case of a double-line spectroscopic binary.

The temporal analysis of our blue spectral series was performed using the generalized Fourier method derived by \citet{HMM}, and revised by \citet{gosset30a}. This technique is especially adapted to the case of unequally spaced data. This method has been applied to the radial velocity time series (see Section\,\ref{rvts}), and to line profiles (see Section\,\ref{lpv}). In the latter case, our approach consisted in computing the power spectrum at each wavelength step leading to a two-dimensioned power spectrum. Mean periodograms were then computed to obtain an overview of the power as a function of frequency for the whole wavelength interval considered in the temporal analysis. We also applied prewhitening techniques considering candidate frequencies in order to check their capability to reproduce the spectral variations \citep[see e.g.][for applications of these techniques to line profile variability analyses]{oef2}.

\subsection{Radial velocity time series \label{rvts}} 

We measured radial velocities on most of the prominent absorption lines present in our blue spectra (see upper panel in Figure\,\ref{spectvs}), namely He\,{\sc i} $\lambda$ 4471, Si\,{\sc iii} $\lambda\lambda$ 4552,4567, C\,{\sc iii} $\lambda\lambda$ 4647,4650, and He\,{\sc i} $\lambda$ 4713, by fitting Gaussians. We applied the generalized Fourier method to radial velocity time series for each individual line, and for the mean of the radial velocities obtained for all these lines (see Table\,\ref{log}). The power spectra are clearly dominated by two peaks, a first one at a frequency ($\nu_1$) of 0.1975\,d$^{-1}$ and the second one ($\nu_2$) close to 0.05\,d$^{-1}$ (along with their $1 - \nu$ aliases). Our time series allowed us to investigate frequencies up to about 10\,d$^{-1}$. However, as the strongest peaks were located in the frequency interval between 0 and 1\,d$^{-1}$, we will restrict the discussion to time scales larger than 1\,d. The lack of significant power at frequencies larger than 1\,d$^{-1}$ is in agreement with the absence of detected variations in time series made of spectra collected during a same night (see the journal of observations in Table\,\ref{log} for details on the time coverage of our data). A summary of the main frequencies reported for the radial velocity time series is given in Table\,\ref{rvnu}. These two frequencies correspond respectively to time-scales of 5.063 and about 20\,d (respectively P$_1$ and P$_2$ hereafter). We also applied the simultaneous search for periods method used for instance by \citet{gosset30a}. This method constitutes some kind of higher order Fourier technique, taking into account the existing correlations between different frequencies. As a result, we found again the two frequencies $\nu_1$ and $\nu_2$ reported above. We note that the latter approach seems to priviledge a value of 0.055\,d$^{-1}$ for $\nu_2$ even for the mean radial velocity time series, although the sequential frequency removal approach pointed to slightly different values.

\begin{figure}
\centering
\includegraphics[width=80mm]{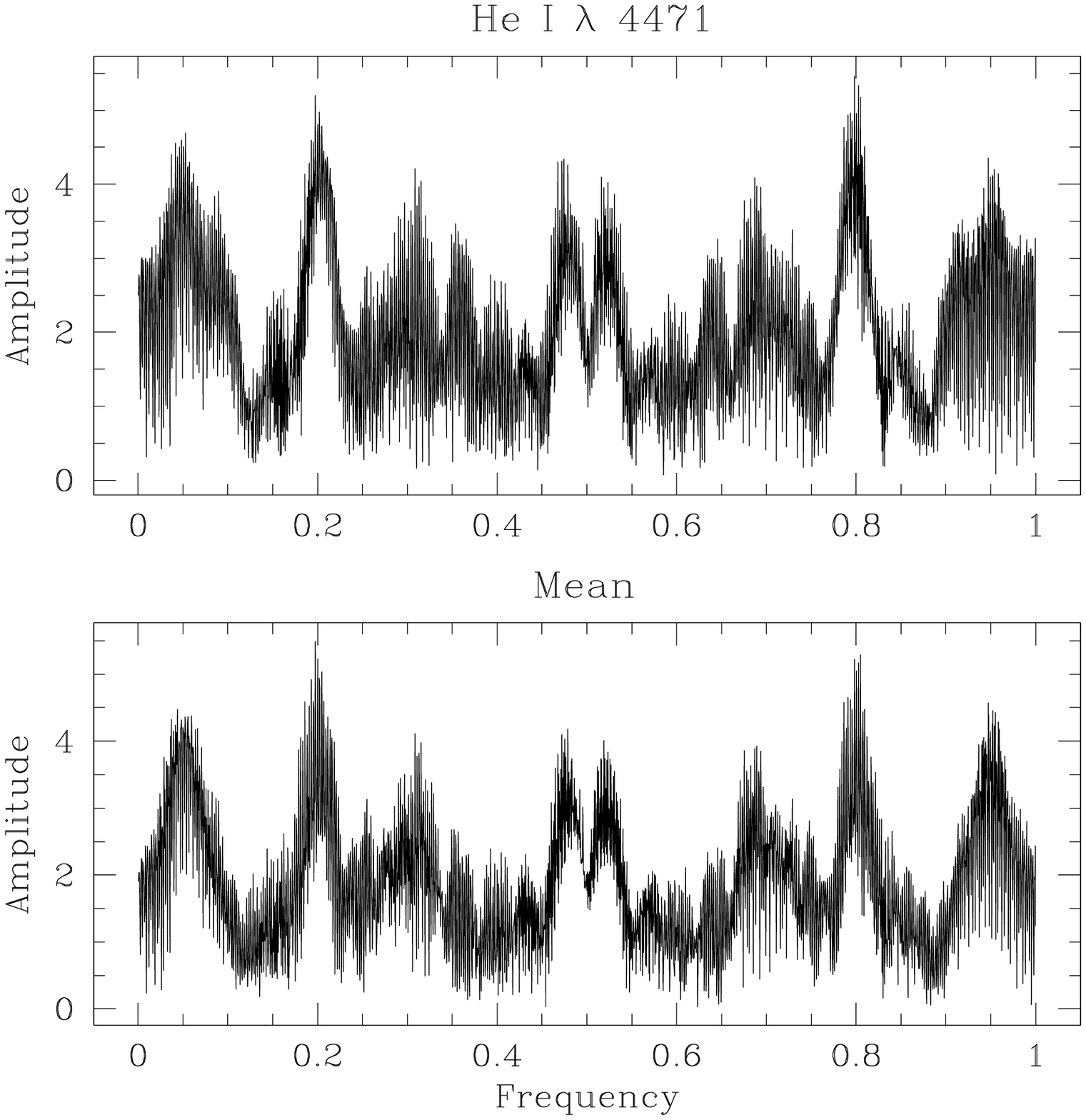}
\caption{Power spectra of the He\,{\sc i} $\lambda$ 4471 (upper panel) and mean (lower panel) radial velocity time series. Both power spectra are dominated by $\nu_1$ and $\nu_2$, along with their $1-\nu$ aliases. The frequency is expressed in d$^{-1}$ and the amplitude in km\,s$^{-1}$.}
\label{rvfour}
\end{figure}

In order to check for the plausibility of these frequencies, we randomly removed about 25\,$\%$ of the data from our radial velocity time series and we repeated the Fourier analysis. We found that the main peaks corresponding to $\nu_1$ and $\nu_2$ are still clearly present in the power spectrum. The latter result lends strong support to the idea that the presence of these peaks is not related to a particular realization of a stochastic process unrelated to an actual physical behaviour. As the simultaneous period removal approach is more reliable than the sequential one for the study of multiperiodic processes, we will favor $\nu_2$\,=\,0.0550\,d$^{-1}$ (corresponding to P$_2$\,=\,18.182\,d) in the following discussion, even though this value deviates significantly from those quoted in Table\,\ref{rvnu}.

\begin{table}
\begin{center}
\caption{Main frequencies expressed in d$^{-1}$ derived from the Fourier analysis of the radial velocity time series. The frequencies marked with a $^*$ correspond to peaks slightly weaker than that of their $1 - \nu$ alias.\label{rvnu}}
\begin{tabular}{l c c}
\hline\hline
Spectral line & $\nu_1$ & $\nu_2$ \\
  & (d$^{-1}$) & (d$^{-1}$) \\
\hline
\vspace*{-0.2cm}\\
He\,{\sc i} $\lambda$ 4471 & 0.1975$^*$ & 0.0415 \\
Si\,{\sc iii} $\lambda$ 4552 & 0.1975 & 0.0485 \\
Si\,{\sc iii} $\lambda$ 4567 & 0.1975 & 0.0440 \\
C\,{\sc iii} $\lambda$ 4647 & 0.1975 & 0.0475 \\
C\,{\sc iii} $\lambda$ 4650 & 0.1975 & 0.0475 \\
He\,{\sc i} $\lambda$ 4713 & 0.1975 & 0.0415 \\
Mean & 0.1975 & 0.0440$^*$ \\
\vspace*{-0.2cm}\\
\hline
\end{tabular}
\end{center}
\end{table}

We fitted sine functions with the period P$_1$ to our radial velocity time series. We then subtracted the fitted radial velocities from the measured ones, and we folded the residual radial velocities with the period P$_2$. We then proceeded the same way but by fitting the measured radial velocities with a period P$_2$ sine function, and folding the residual radial velocities with P$_1$. We repeated the same procedure for the measurements obtained from every line, and for the mean of all lines used for the measurements. The resulting residual radial velocity curves suggest trends a priori not compatible with noise. Even though the semi-amplitude of the radial velocity variations are of the same order of magnitude than the expected typical error on our measurements\footnote{We estimate the peak-to-peak amplitude for the variations of the radial velocities of Diffuse Interstellar Bands (DIBs) to be of the order of 6\,km\,s$^{-1}$.}, these curves suggest that two simultaneous periodic phenomena are responsible for the variations revealed by the temporal analysis. The consistency of the results obtained for the different lines concerning $\nu_1$ indicates that its presence is firmly established by our temporal analysis. However, the deviations observed from one line to the other for $\nu_2$ deserve some caution concerning the corresponding variability time scale. The different values close to $\nu_2$ reported in Table\,\ref{rvnu} allow a time scale ranging between about 20 and 24 days, down to 18 days according to the results of the simultaneous search for periods.

\subsection{Line profile variability \label{lpv}}

\begin{figure}
\centering
\includegraphics[width=80mm]{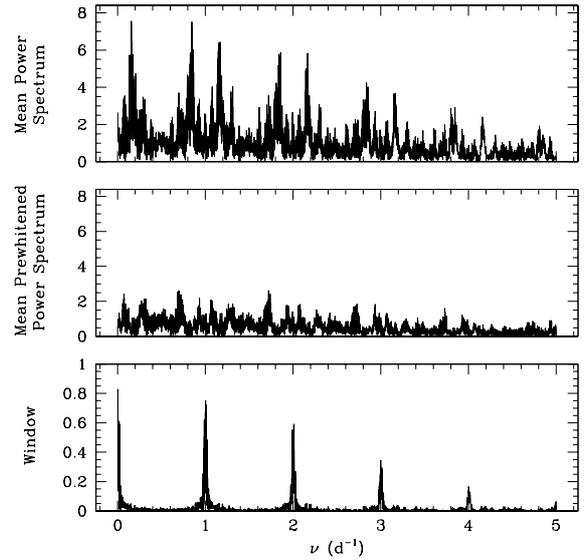}
\caption{Fourier analysis of the H\,$\beta$ line betwen 4860 and 4863\,\AA\,. {\it Upper panel:} mean power spectrum between 0 and 5\,d$^{-1}$. {\it Middle panel:} mean power spectrum obtained after prewhitening using the frequency of the highest peaks of the initial power spectrum, i.e. 0.1975 and 0.15375\,d$^{-1}$. {\it Lower panel:} spectral window directly related to the sampling of the time series.}
\label{hbetafour}
\end{figure}

Beside the radial velocity variations discussed above, we also detect a substantial line profile variability in H\,$\beta$. Our temporal analysis using the same tools as \citet{oef2} reveals a variability dominated by a frequency of 0.15625\,d$^{-1}$ ($\nu_3$, corresponding to a time scale close to 6.4\,d, hereafter P$_3$). After prewhitening with this frequency, the residual power spectrum is dominated by $\nu_1$. When the power spectrum is prewhitened using $\nu_1$, $\nu_3$ is slightly shifted to a lower value, i.e. 0.15375\,d$^{-1}$. The resulting periodogram after subtraction of the two reported time scales is presented in Fig.\,\ref{hbetafour}. Considering the uncertainties on the frequencies emerging from our temporal analysis, $\nu_3$ is compatible with the difference $\nu_2 - \nu_1$ (mostly if $\nu_2$ is close to 0.044\,d$^{-1}$. In that case, the variations in the line profile of H\,$\beta$ would therefore be totally determined by the frequencies $\nu_1$ and $\nu_2$. However, the prewhitening using $\nu_1$ and $\nu_2$ gives a residual power spectrum dominated by a peak close to $\nu_3$, with a varying amplitude depending on the chosen value for $\nu_2$. Alternatively, $\nu_3$ may thus be related to an independent process. 

The Fourier analysis of the He\,{\sc i}, Si\,{\sc iii} and C\,{\sc iii} line profiles reveals also the presence of strong peaks at frequencies at -- or very close to -- $\nu_1$ and $\nu_2$. We report also on the presence of these frequencies in the central part of He\,{\sc ii} $\lambda$ 4686 (displaying a weak central emission surrounded by two shallow absorption parts), with a weak residual peak at a frequency close to $\nu_1 + \nu_2$. For the red absorption part of the same line, the situation is more confused but $\nu_1$ is present.

\begin{figure}
\centering
\includegraphics[width=80mm]{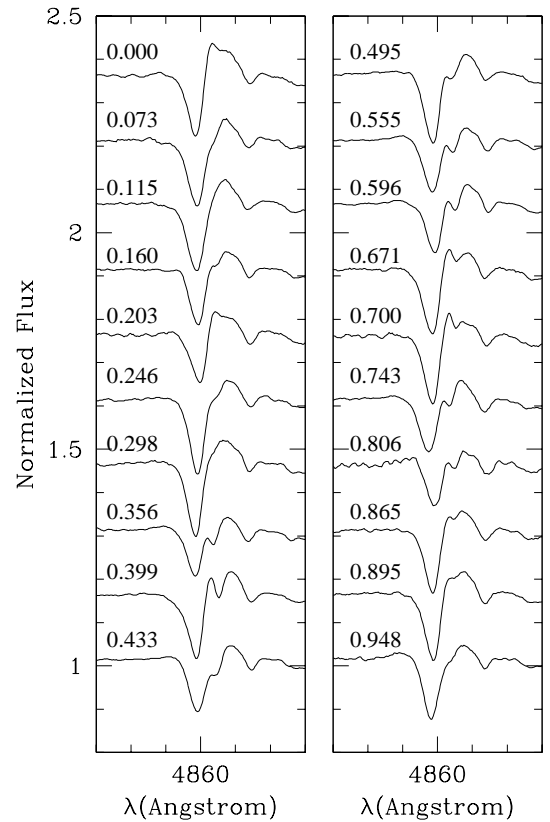}
\caption{Plot of a sample of 20 H\,$\beta$ profiles (between 4845 and 4870\,\AA\,) folded in phase according to the time scale P$_1$. The phase is specified in each case. The T$_\circ$ was arbitrarily chosen as the time of the first spectrum of the time series. For the sake of clarity, the plots are shifted in normalized flux by an arbitrary value of 0.15 units. The spectra have been selected in order to reach a quasi uniform phase coverage of the candidate period P$_1$. The apparently constant absorption feature at about 4867\AA\, is attributed to N\,{\sc iii}.}
\label{plothbeta}
\end{figure}

H\,$\beta$ is the only line in the blue spectrum of HD\,195592 that presents a strong line profile variability. To illustrate it, we have plotted a sample of the profiles in Figure\,\ref{plothbeta}. In order to check for any trend related to $\nu_1$, i.e. the most clearly present frequency arising from our temporal analysis, we have folded the profiles using the corresponding period P$_1$. The selected spectra have been chosen in order to sample as uniformly as possible the complete phase interval between 0 and 1. Even though some consecutive profiles display obvious similarities in the plotted sequence, therefore suggesting a progressive and continuous profile variation, it should also be noted that anomalies are observed as compared to an expected smooth variation ruled by a unique and well determined time scale. This may be due to the simultaneous impact of P$_2$ on the line profile variations.

As in the case of the radial velocity time series, $\nu_1$ appears much more firmly established than $\nu_2$, as it is found invariably in all lines investigated in the blue spectrum, with no deviation with respect to the value given in Table\,\ref{rvnu}.

\section{Discussion} \label{disc}

The main result reported above is the signature of a spectroscpic binary with at least two simultaneous variability time-scales for the radial velocities. Considering the apparent luminosity ratio suggested by the two blended components in the right part of Figure\,\ref{specred}, the secondary might be a B-type star. As the SB2 signature is found in only one line among those accessible in our data, it is not possible to speculate on the accurate spectral classification of the secondary. Both spectral types and luminosity classes affect the luminosity ratios of the two stars of the system, and it is not possible to translate an equivalent width ratio of one single line of the two components into a spectral classification for both stars. The fact that asymmetric profiles are not obvious in the blue domain points to low amplitudes radial velocity excursions (in agreement with the measurements discussed in Section\,\ref{rvts}). In the context of a binary scenario, one of the two time-scales (P$_1$ or P$_2$) should be the orbital period. The fact that P$_1$ is much more firmly established than P$_2$ leads us to consider the former as the better candidate for the binary period. However, the origin of the second time-scale needs to be investigated. Potential scenarios are considered below.

\subsection{The triple system scenario}
A biperiodicity in the radial velocity variations is reminiscent of what one may expect from a triple system. In the context of this scenario, we can envisage two separate situations:
\begin{enumerate}
\item[] Case 1: only the primary spectrum is detected because the lines of the other two stars are too weak, or even absent. In this case, the observed radial velocity time series is the result of two superimposed orbital motions.
\item[] Case 2: the observed line profiles are the result of the blend of two or three components associated with the stars of the multiple system.  
\end{enumerate} 

\subsubsection{Case 1}
As a first approximation, we may consider that the observed radial velocity variations are due to one of the two orbits, perturbed by the second one. The measured radial velocities are therefore representative of the motion of the primary star, and Kepler's laws should apply.

We followed an iterative orbital disentangling procedure in order to separate the orbital elements of the two putative orbits. Let us define the following quantities, where $i$ goes from 1 to $N$\footnote{In the present case, N is equal to 81, the total number of spectra in our sample.}:
\begin{enumerate}
\item[] $D_i$ : radial velocity measurements
\item[] $R_{i,j}$ : theoretical radial velocities of Orbit 1 at iteration~$j$
\item[] $Q_{i,j}$ : theoretical radial velocities of Orbit 2 at iteration~$j$
\item[] $X_{i,j} = D_i - Q_{i,j}$ : radial velocities corrected for Orbit 2 at iteration $j$, i.e. approached radial velocities of Orbit~1
\item[] $Y_{i,j} = D_i - R_{i,j}$ : radial velocities corrected for Orbit 1 at iteration $j$, i.e. approached radial velocities of Orbit~2
\end{enumerate}

As a starting point of the procedure, we estimated the $R_{i,0}$ on the basis of a simple sine function with a period equal to P$_1$. We then calculated $Y_{i,1}$, and we computed the first approached orbital elements of Orbit 2, using the Liege Orbital Solution Package (LOSP, see \citet{losp}), as already applied by \citet{ic1805_2}. On the basis of the latter orbital solution, we derived $Q_{i,1}$ and the corresponding $X_{i,1}$. We used LOSP iteratively and convergence was generally obtained for $j$ = 2 or 3.

The semi-amplitude of the radial velocity curves of the two putative orbits computed by LOSP are respectively equal to 5.1\,$\pm$\,0.8 and 5.5\,$\pm$\,0.8\,km\,s$^{-1}$. In addition, the two orbits are compatible with null or small eccentricities depending on the value adopted for P$_2$. The low amplitude of the radial velocity variations strongly limits the quality of the results, and the orbital elements are computed from radial velocity series which are corrected on the basis of approximated theoretical radial velocities carrying inevitably additional uncertainties. The parameters we derived request therefore some caution. Finally, we note that we obtained a very low value for the projected semi-major axis : of the order of 0.5 and 2.0\,R$_\odot$ respectively for the two orbits. If real, these values either indicate that the inclinations of the orbits are very low (i.e. the system is viewed pole-on), or that the secondary and tertiary components have very low masses compared to the primary.

\subsubsection{Case 2}
When line profiles are made of two blended components respectively related to two orbiting stars, a slight variation in the centroid of a spectral line is expected. In such a situation, the measured radial velocities are therefore not a measurement of the orbital motion of one star, but would rather represent the variations of the centroid of a spectral line made of at least two slightly shifted components. In this context, as the components are not disentangled, the measured radial velocity variations are not obeying Kepler's laws, and the orbital elements derived above in Case 1 do not make sense anymore. Considering the significant detection of a secondary component at least in  He\,{\sc i} $\lambda$ 6678, this scenario seems indeed plausible.

We had a very close look at line profiles (mainly He\,{\sc i} $\lambda$ 4471, the strongest absorption line in the blue spectrum) in order to seek for asymmetries, and slight anticorrelated width and depth variations. We detect marginal variations of the depth of the line, and slight asymmetries with the red wing of the absorption profile steeper than the blue one. However, such an asymmetric profile may result from a weak emission component in the red wing of the He\,{\sc i} line related to the wind material, and is therefore not necessarily due to a blue-shifted secondary absorption component. 

If the second time-scale P$_2$ is indeed related to the presence of a third -- still undetected -- star orbiting along a wider orbit, the measured radial velocity variations could be explained by the modulation of the blended profile of the central binary (discussed a few lines above) by the motion of the putative tertiary star. In the context of such a scenario, we would be dealing with a slightly hierarchical triple system with a rather low period ratio P$_2$/P$_1$ ($\sim$\,4). Considering a mass of about 30\,M$_\odot$ for the O9.5I primary \citep{martins}, and assuming a mass of about 20\,M$_\odot$ for the secondary, we applied the stability criterion established by \citet{leonardduncan} for hierarchical triple systems\footnote{As our data did not allow to determine the spectral type of the secondary, our assumption relied on the fact that the secondary is significantly less luminous than the primary, but still detected. We note however that the value of the inner mass ratio does not have a strong impact on the present discussion.}. Our result is that such a system may be dynamically stable provided the mass of the third star is not larger than about 7\,M$_\odot$. This upper limit on the third mass translates into a spectral type not hotter than a late-B if it is a main-sequence star, or even late-A if it is a supergiant. Consequently, even though the period ratio of such a hypothetical hierachical system is quite small, its existence should not completely be ruled out provided the mass of the third star is low enough.

\subsection{Stellar pulsations}

Line profile deformations responsible for a slight shift of the centroid of a line profile can also be caused by pulsations of a single star. However, such phenomena are expected to occur on time scales of a few hours \citep[see e.g.][]{kambe-zetaoph,hd93521,ibvsnrp}, much shorter than the time scales reported above in the case of HD\,195592. Non-radial pulsations may give rise to frequencies of a fraction of a day only if beating or resonance is occurring \citep{kaufermultiper}, but this is a priori unlikely to happen without the presence of parent peaks at much higher frequencies. We insist on the fact that the power decreases rapidly as a function of frequency in our power spectra, lending strong support to the idea that the variability time scales are of several days. We therefore do not consider non-radial pulsations as a plausible explanation for the variations revealed by our spectral time series.\\

\subsection{Stellar rotation}

On the other hand, one may also consider one of the reported variability time-scales to be the rotation period. We used the relations (1 to 3) from \citet{oef2} in order to estimate the expected minimum (constrained by the critical rotation velocity) and maximum (constrained by the projected rotational velocity) rotation periods. To do so, we considered stellar parameters as prescribed by \citet{martins} for an O9.5 supergiant (M$_*$\,=\,30\,M$_\odot$, R$_*$\,=\,23\,R$_\odot$ and L$_*$\,=\,3.1\,$\times$\,10$^{5}$\,L$_\odot$), along with the projected rotational velocity determined by \citet{CE}, namely 60\,km\,s$^{-1}$. These values lead to minimum and maximum rotation periods respectively of 2.7 and 19.4\,d. P$_1$, P$_3$ and to some extent P$_2$ are therefore compatible with the allowed interval. Therefore, we cannot reject a scenario where one of the time-scales might be related to the stellar rotation. For instance, the presence of P$_3$ in H\,$\beta$ may point to a rotational modulation of the profile produced partly in the wind, with $\nu_2$ resulting from a beating between $\nu_1$ and $\nu_3$. However, if $\nu_3$ is indeed a true frequency, we would have to explain why it is absent from other lines although $\nu_2$ is ubiquitous in absorption lines. Alternatively, P$_3$ in H\,$\beta$ may still be related to the stellar rotation of the primarywhatever the origin of P$_2$.

\subsection{HD\,195592 as a runaway \label{runaway}}

HD\,195592 has been classified by \citet{mcswainrunaway} as a runaway on the basis of a peculiar space velocity ($v_{pec}$) criterion, i.e. $|v_{pec}|\,>\,30 + \sigma_{vpec}$ where $\sigma_{vpec}$ is the error on $v_{pec}$. They determined a peculiar space velocity of --42.3\,$\pm$\,8.2\,km\,s$^{-1}$. This value was calculated using radial velocities similar to ours. The runaway classification proposed by \citet{mcswainrunaway} is therefore not contradicted by our measurements. The parent cluster of HD\,195592 remains unknown, even though the study of \citet{dewit-fieldO} revealed the presence of a nearby cluster.

As mentioned in Section\,\ref{intro}, several runaways have already been identified as binaries harbouring a compact companion. For instance, \citet{mcswainrunaway} summarize the orbital elements of five spectroscopic binaries that belong to the category of runaways. In the case of HD\,195592, we report above on the detection of a secondary that may be a B-type star. Even though its spectral type is not determined, the presence of its spectral signature rules out a degenerate secondary star. We may envisage two situations:
\begin{enumerate}
\item[-] HD\,195592 is an SB2 system, with the second variability time-scale being related to the stellar rotation. Such a binary system may have been ejected through dynamical interactions in its parent open cluster.
\item[-] HD\,195592 is a triple system, made of a close massive binary plus a much lower mass unidentified third star. In such a situation, we would be dealing with the only runaway triple system identified so far. However, our analysis did not provide clear evidence for the presence of a third star. The triple system scenario should therefore only be considered as a working hypothesis for future investigations devoted to HD\,195592. Considering the low amplitude variations of the radial velocities in this system, a much higher spectral resolution is needed in order to (i) confirm the presence of a third star and (ii) identify its nature\footnote{At this stage, it it interesting to note that our data cannot reject a triple scenario including a neutron star as a third object. Such a situation may occur in the context of the supernova ejection processes, even though it is not clear so far whether such a multiple system could survive to the supernova explosion. In addition, the presence of the neutron star on the wider orbit of the hierarchical triple system would deserve some attention, as a mass segregation is generally observed in high multiplicity system. It is so far not sure that dynamical interactions consecutive to the supernova explosion could lead to some orbital exchange within the system.}. Such an identification is requested if one wants to better understand the nature of HD\,195592 in the framework of the study of runaway stars.
\end{enumerate}

\section{Summary and conclusions\label{conc}} 
We found strong evidence for the signature of a spectroscopic binary in the He\,{\sc i}\,$\lambda$\,6678 line of the runaway supergiant HD\,195592. Using unprecedented spectral time series, we investigated the radial velocities and the line profile variability in the blue part of the optical spectrum. We report on a low amplitude variability of the radial velocities with frequencies $\nu_1$ and $\nu_2$ respectively of 0.1975\,d$^{-1}$ (P$_1$\,=\,5.063\,d), and close to 0.05\,d$^{-1}$ (P$_2$\,$\sim$\,20\,d), in all prominent absorption lines. We also performed a line profile variability analysis of the H\,$\beta$ line which displays complex variations, dominated by a frequency close to 0.15\,d$^{-1}$, in addition to $\nu_1$ that is also clearly present.

We attribute the period P$_1$ to the orbit of a close binary system made of an O9.7I primary with a probable B companion. In this scenario, the line profile variations of H\,$\beta$ may be influenced by the combined effect of the binary motion and of an intrinsic variability of the primary. The second time-scale (P$_2$) may be the signature of the presence of a putative third star moving along a somewhat wider orbit. Dynamical stability considerations suggest a triple system with such a low period ratio may exist provided the mass of the third star is low enough. Strong evidence for the presence of a third star is however still lacking. Alternatively, the second time-scale may be related to variations intrinsic to the stellar wind of the supergiant primary, possibly due to the stellar rotation. We note that in the context of the investigation of runaway stars, the case of HD\,195592 deserves a particular attention in the sense that it is now confirmed as a SB2 system, in contrast with other runaway binaries identified as SB1 systems such as HD\,14633 and HD\,15137.

The results of the present study allow us to draw guidelines for future observations. In order to clarify the nature of HD\,195592, the signature of the SB2 system should be searched for in the complete visible spectrum, investigating as many lines as possible, and the putative signature of the third star should be scrutinized. In addition, observations in the X-rays may provide valuable information as well. HD\,195592 is included in the field of two {\it ROSAT} observations (Seq. IDs RS930941N00 and RS930942N00) but does not seem to be a bright soft X-ray emitter. However, more sensitive observations covering a wider energy doamin (e.g. with {\it XMM-Newton}) are expected to be helpful. 

\section*{Acknowledgements}
We would like to express our gratitude to Dr Hugues Sana for preparing and providing LOSP, and to Dr Eric Gosset for stimulating discussions. The travels to OHP were supported by the Minist\`ere de l'Enseignement Sup\'erieur et de la Recherche de la Communaut\'e Fran\c{c}aise. This research is also supported in part through the PRODEX XMM/Integral contract, and more recently by a contract by the Communaut\'e Fran\c{c}ais de Belgique (Actions de Recherche Concert\'ees) -- Acad\'emie Wallonie-Europe. We would like to thank the staff of the Observatoire de Haute Provence (France) for the technical support during the various observing runs. The SIMBAD database has been consulted for the bibliography.

\end{document}